# Multilayer graphene condenser microphone

*Dejan Todorović[1,2], Aleksandar Matković[3], Marijana Milićević[3], Đorđe Jovanović[3], Radoš Gajić[3], Iva Salom[4] and Marko Spasenović[3]*

[1]School of Electrical Engineering, University of Belgrade, Bulevar kralja Aleksandra 73, 11120 Belgrade, Serbia

[2]Dirigent Acoustics Ltd, Mažuranićeva 29/9, 11050 Belgrade, Serbia

[3]Center for Solid State Physics and New Materials, Institute of Physics Belgrade, University of Belgrade, Pregrevica 118, 11080 Belgrade, Serbia

[4]Institute Mihailo Pupin, University of Belgrade, Volgina 15, 11060 Belgrade, Serbia

ABSTRACT: Vibrating membranes are the cornerstone of acoustic technology, forming the backbone of modern loudspeakers and microphones. Acoustic performance of condenser microphone is derived mainly from the membrane's size and achievable static tension. The widely studied and available nickel has been the one of dominant membrane material for several decades. In this paper we introduce multilayer graphene as membrane material for a condenser microphone. The graphene device outperforms a high end commercial nickel-based microphone over a significant part of the acoustic spectrum, with a larger than 10 dB enhancement of sensitivity. Our experimental results are supported with numerical simulations, which show that a 300 layer thick graphene membrane under maximum tension would offer excellent extension of the frequency range, up to 1 MHz, with similar sensitivity as commercial condenser microphones.

1. Introduction

Graphene has emerged as an exciting new material for fundamental research and novel applications, due to its many remarkable properties such as ultrahigh carrier mobility [1], high optical transparency [2,3], and enormous chemical reactivity [4]. Furthermore, graphene shows record high thermal conductivity [5] and Young's modulus [6], making it an excellent thin membrane. Measurements have shown that monolayer graphene has a breaking strength of 42 N/m which is over 200 times greater than a hypothetical steel film of the same thickness, with a tensile modulus of 1 TPa [6]. Graphene membranes are lightweight, and their mechanical, thermal and electrical properties promise remarkable possibilities in acoustic applications. Although electrical and optical properties of graphene have been extensively studied and applied, the first utilizations of the mechanical properties of graphene in acoustics have only recently been realized, with the first demonstrations of a graphene thermoacoustic device [7-9], electrostatic loudspeaker [10] and a graphene earphone [11,12].

Vibrating lightweight membranes are of tremendous relevance in the acoustic industry, as exemplified by the reign of condenser microphones as the best platform for recording and measuring acoustic waves using the same pressure detection principle as the human ear, with high quality, repeatable and predictable performance [13]. Condenser microphones, discovered at Bell Labs in 1916, utilize a vibrating membrane as one plate of a capacitor charged by a polarization voltage. Incident acoustic waves set the membrane in motion, which is detected with an attached electronic circuit. Most improvements to condenser microphones in the last three decades focused on the electronics, for example on improving the analogue to digital converters and placing them inside the microphone body to minimize interference and noise in analogue signal transmission chain. An exception is the rise of microphones based on microelectromechanical systems (MEMS), typically using thin silicon as the membrane material [14]. MEMS-based microphones offer the advantage of small size, which is important for applications in hearing aids and aeroacoustic arrays [15]. MEMS microphones feature membranes with a thickness on the order of 1 $\mu$m and have shown a frequency response similar to conventional macroscopic microphones [16]. The nanoscale thickness of graphene membranes and their exceptional mechanical properties make them an excellent candidate for even smaller microphones with enhanced spectral response.

Here we report an experimental realization of a standard condenser microphone with a multilayer graphene membrane with performance comparable to a professional microphone, at a membrane thickness of only 25 nm. Packaged in the same commercial casing and using the same electronic amplifier, the graphene microphone sensitivity outperforms a professional microphone (Brüel & Kjaer (B&K) 4134 microphone) by 12 dB, at frequencies up to 12 kHz. The experimental data are supported with numerical simulations, which additionally show that a membrane consisting of 300 layers of graphene would respond in a usable frequency range extending up to 1 MHz, in the ultrasonic part of the spectrum.

2. Fabrication

Our sample consists of multilayer graphene grown on nickel foil with chemical vapour deposition (Graphene Platform). The nickel is etched away in a 40 mg/ml iron chloride water solution, yielding a floating multilayer graphene film (Figure 1a). The graphene film is scooped out of the solution onto a supporting polyethylene terephthalate (PET) frame which has a circular hole in the center (Figure 1b). The graphene covers the hole to form the membrane. The diameter of the hole and membrane was varied between 5 mm and 12 mm with similar results. The membrane is subsequently left to dry for 24 hours prior to mounting the device into the microphone casing. This fabrication procedure results in a high yield of membranes, at a success rate larger than 70%. To increase the yield, the membranes were not rinsed in water or other solvents [6]. As a result, a residue of iron chloride is left on the membrane, as confirmed with scanning electron microscopy (see Supplementary Information). The residue does not appear to significantly affect the performance of the microphone, as demonstrated in the following paragraphs. Raman spectroscopy was used to confirm that the membrane consists of graphene (Supplement). Atomic force microscopy of sacrificial membranes indicated an average thickness of 25 nm (Supplement).

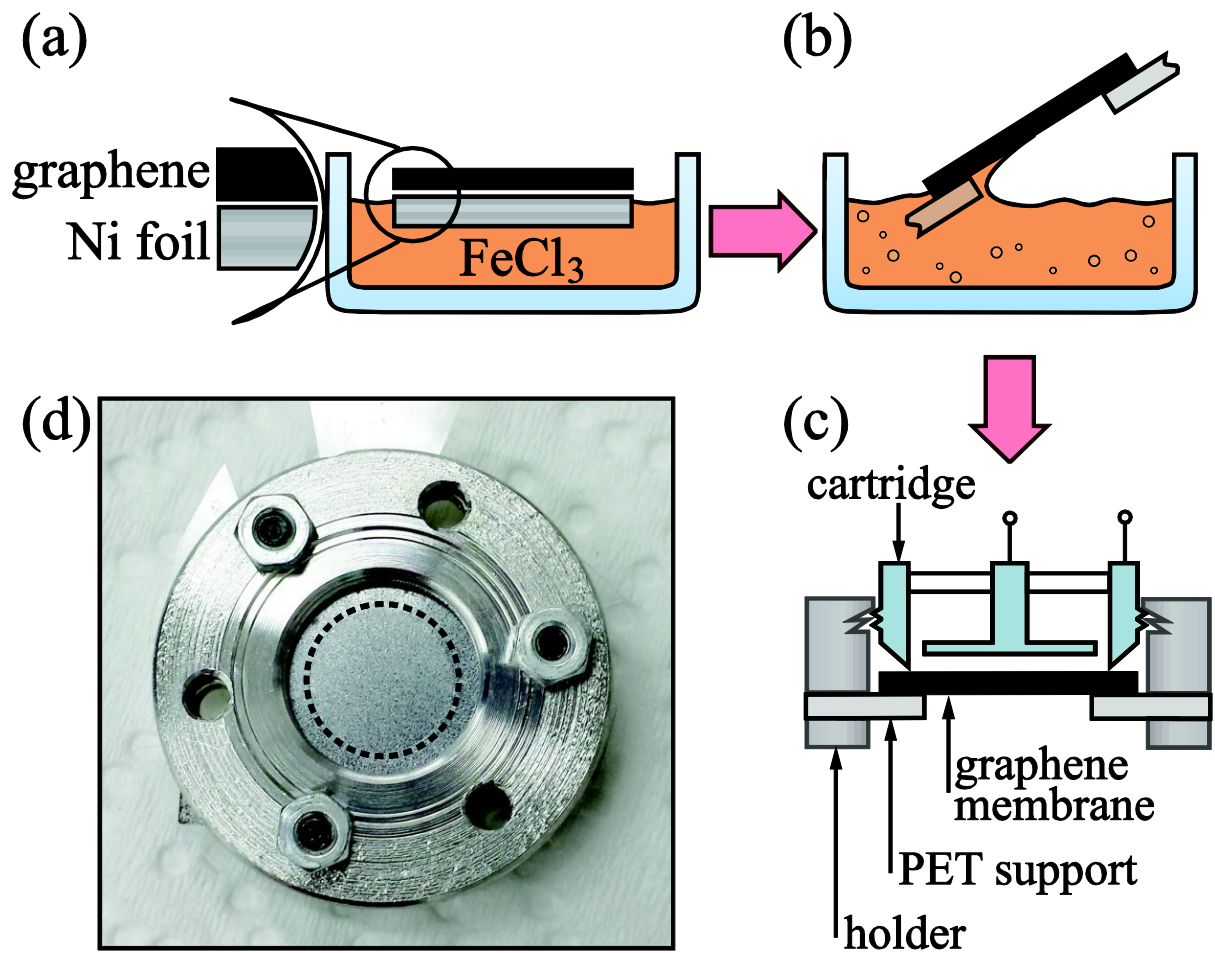

**Figure 1.** Preparation of the graphene membrane. Multilayer graphene on nickel is placed in an etching solution (a). The floating graphene layer is scooped onto a support with a predefined hole for the membrane (b). The membrane is loaded onto a holder which is then fixed into the professional microphone cartridge (d). A photograph of the graphene membrane in the microphone holder (d). The dashed line encircles the freestanding part of the membrane.

3. Measurements

Microphone performance is measured by recording a constant amplitude sine sweep from 10 Hz to 24 kHz, played on a small loudspeaker. The signal was amplified with a professional microphone preamplifier (B&K ZC0032) and analyzed with a PrismSound dScope Series III signal generator and analyzer. The comparison (results in Figure 2.) and substitution (result in Figure 4.) methods were used to test several graphene microphone against the professional microphone (details are given in the Supplementary Information). During the measurements a DC polarization voltage of 200 V was applied to the microphones. Frequency response of small loudspeaker at 90° incidence with professional and graphene microphone is presented in Figure 2.

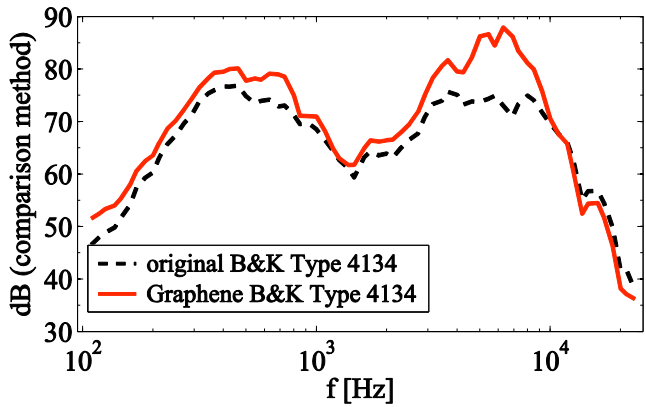

**Figure 2.** Frequency response of the small loudspeaker measured with a 5 mm membrane diameter graphene microphone (red curve) and a chosen professional microphone (black dashed curve), using the comparison method at 90° incidence angle. Sensitivity is higher than the professional microphone at all frequencies below 11 kHz.

The distance between the membrane and the oppositely charged condenser back plate was crudely controlled by tightening and loosening the holder screw (see Supplementary Information). The microphone sensitivity increases as the membrane approaches the back plate, down to a minimal distance of 18.6 μm allowed by the microphone casing. The membrane static tension is estimated using the membrane collapse effect [18,19] as:

$$T_{m0} = \sigma t = \frac{27\varepsilon_0 r^2 V_p^2}{64 h^3} \left[\frac{N}{m}\right],$$

where $\sigma$ is membrane tensile stress (N/m²), $t$ is membrane thickness (m), $\varepsilon_0$ is dielectric permittivity of air (= 8.85x10⁻¹² F/m), $r$ is radius of membrane (m), $V_p$ is polarization voltage (V) and $h$ is nominal distance from diaphragm to back plate (m).

For a polarization voltage of 200 V and a membrane radius of 5 mm, the estimated static tension is approximately 640 N/m.

The static tension on the graphene membrane is five times smaller than the manufacturer specified tension on the nickel membrane of the professional microphone. A small static tension results in a large microphone efficiency, as in [20]:

$$M = \frac{V_p r^2}{8 h T_{m0}} \left[\frac{\text{V}}{\text{Pa}}\right],$$

and is hence the most likely reason for the superior performance of the graphene microphone. The tradeoff of small membrane tension is a low cutoff frequency, given as [21]:

$$f_{max} = \frac{2.4}{2\pi r} \sqrt{\frac{T_{m0}}{\sigma_m}},$$

where $\sigma_m$ is surface mass density of membrane (kg/m²).

However, the graphene membrane has a surface mass density three orders of magnitude smaller than that of the nickel membrane of the professional microphone, which compensates for the weak tension.

4. FEM simulation and measurements

To reinforce the experimental results and explore the limits of graphene microphone membranes, we employ finite element method (FEM) calculations. We start from the well-known publicly available professional condenser microphone model in Comsol Multiphysics.

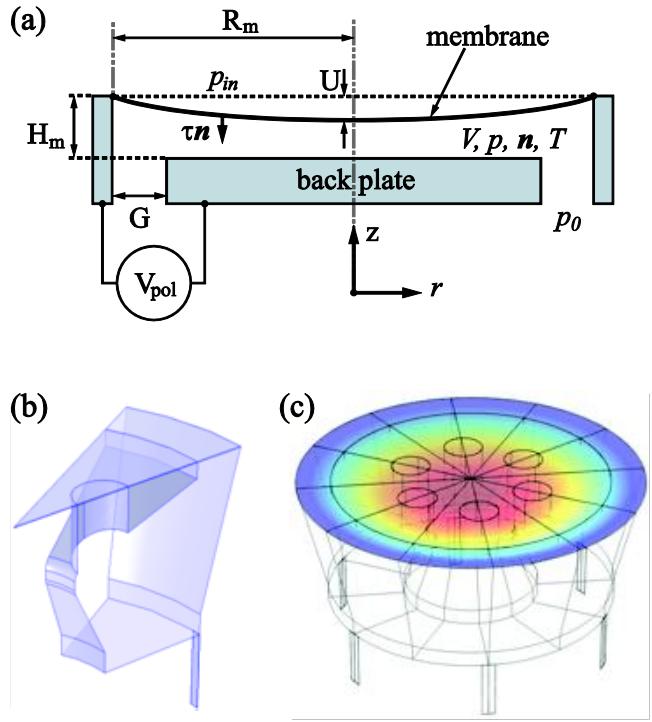

**Figure 3.** Model used in FEM simulations. a) 2D condenser microphone simplified model b) Section of the symmetric professional condenser microphone system c) Full 3D model with membrane static deflection (in colour) under applied polarization voltage.

Two different models were employed, a simplified 2D axial model of the condenser microphone (Figure 4.a) and a detailed 3D symmetric model of the professional condenser microphone (Figure 4.b and c). Both models are multiphysics problems that involve several physics interfaces: Thermoacoustics, Electrostatics, Moving Mesh, and a Membrane model. Each finite-element model is solved fully coupled using the frequency-domain linear perturbation solver. This includes the DC charging (pre-polarization) and deformation of the membrane - "pull-in" voltage, which makes out the zeroth order linearization point. Many microphones contain an air-filled back volume below the electrode. In our analysis we disregard the effect of the back pressure of air in this volume. The membrane is deformed due to electrostatic forces from charging the capacitor and the pressure variation due to the incoming uniform acoustic signal pin. The dimensions and parameters of the simulation are given in Table S1 (Supplementary Information). The membrane material is nickel, 5 μm tick, with applied static tension of 3160 N/m.

Simulations were performed for the three cases: the original professional microphone; the same microphone but with a 30-layer graphene membrane replacing the original nickel membrane, as was the case in experiment; and a ¼" microphone with a hypothetical 300-layer graphene membrane (see Supplementary Information). The simulated response spectrum of the 30-layer graphene microphone membrane, compared to the measured microphone response, is depicted in Figure 4. The normalized results, in agreement with the measurements, indicate that the graphene microphone has sensitivity higher than the professional condenser microphone by more than 11 dB, up to a frequency of 11 kHz. At higher frequencies, the larger static tension of the professional microphones nickel membrane starts to dominate, leading to a reduced relative performance of the graphene microphone. The frequency response and sensitivity in the limit of thin membranes are defined by the geometry and membrane static tension, and air load on the membrane dominates the mass of the membrane. In this limit, it is interesting to consider slightly thicker graphene membranes, which could theoretically support a larger static tension. The graphene microphone suffers reduced performance at frequencies larger than 2 kHz, due to the supporting metal ring.

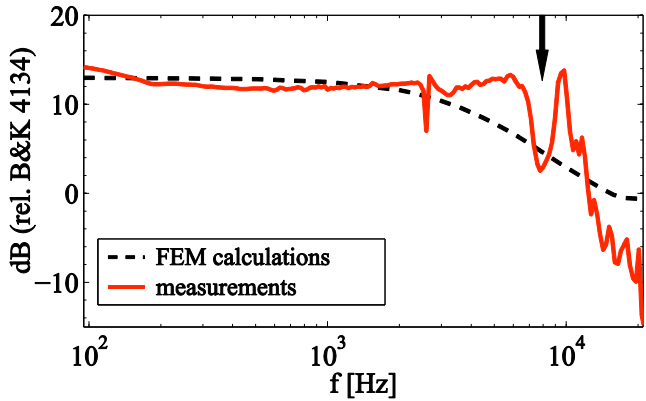

**Figure 4.** Comparison of the measured graphene microphone response (30 layer graphene membrane, red curve) to the FEM calculated response (black dashed curve), all normalized to the average measurements of the original professional condenser microphone. Overall difference in response is an influence of the microphone membrane holder (metal ring: 25 mm outer diameter, 10 mm inner diameter, 3 mm tick).

The strength of a graphene membrane scales linearly with the number of layers [23]. In accordance with Eq. 3, membranes that support a larger static tension have larger cutoff frequencies. Figure 5. depicts simulated response of a 300 layer graphene membrane, under the same experimental conditions achieved in the rest of this work. The microphone exhibits superb performance over a wide frequency range, up to 1 MHz. The microphone sensitivity at frequencies below 1 MHz remains comparable to the professional small membrane professional condenser extended range microphones (B&K 4136 and B&K 4138).

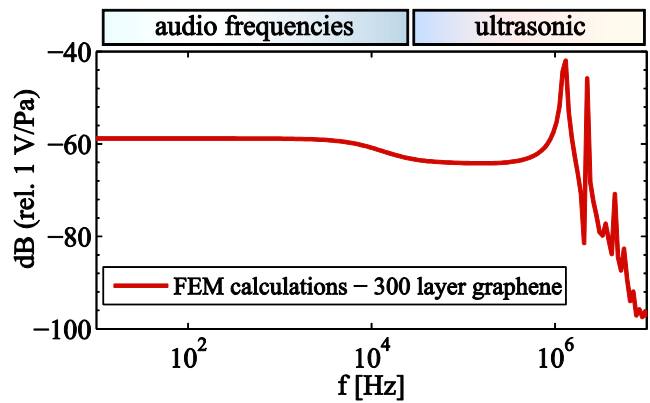

**Figure 5.** FEM simulation of a 300 layer graphene microphone.

5. Conclusions

In conclusion, we demonstrate multilayer graphene microphones with a performance comparable to professional microphones. A microphone with a 30 layer graphene membrane displays up to 15 dB higher sensitivity compared to a commercial microphone, at frequencies up to 11 kHz. Finite element simulations confirm graphene microphone sensitivity and cutoff, and indicate that a microphone with a 300 layer graphene membrane would show similar sensitivity as state of the art condenser microphones at frequencies up to 1 MHz, deeply entering the ultrasonic part of the spectrum. Our work paves the way for the use of widely available and inexpensive graphene in acoustics and touches upon the important ultrasound part of the spectrum, unreachable by the conventional state of the art microphones.

ACKNOWLEDGMENT

This work was supported by Dirigent Acoustics Ltd. and the Serbian Ministry of Education and Science under projects OI171005, III45018 and TR32038.